# A Group based Time Quantum Round Robin Algorithm using Min-Max Spread Measure

Sanjaya Kumar Panda
Department of CSE
NIT, Rourkela

Debasis Dash
Department of CSE
NIT, Rourkela

Jitendra Kumar Rout
Department of CSE
NIT, Rourkela

## ABSTRACT
Round Robin (RR) Scheduling is the basis of time sharing environment. It is the combination of First Come First Served (FCFS) scheduling algorithm and preemption among processes. It is basically used in a time sharing operating system. It switches from one process to another process in a time interval. The time interval or Time Quantum (TQ) is fixed for all available processes. So, the larger process suffers from Context Switches (CS). To increase efficiency, we have to select different TQ for processes. The main objective of RR is to reduce the CS, maximize the utilization of CPU and minimize the turn around and the waiting time. In this paper, we have considered different TQ for a group of processes. It reduces CS as well as enhancing the performance of RR algorithm. TQ can be calculated using min-max dispersion measure. Our experimental analysis shows that Group Based Time Quantum (GBTQ) RR algorithm performs better than existing RR algorithm with respect to Average Turn Around Time (ATAT), Average Waiting Time (AWT) and CS.

## 1. INTRODUCTION
In the time sharing environment, the processes are sharing CPU time one after another. The time is referred as time slice or time interval or TQ. After the specified TQ is expired, the CPU time is used by another process. If a process completes its execution before TQ expired, then next process is assigned to the processor. When TQ is very less, the response time and the context switching are more. Normally, TQ is in between 10 to 100 milliseconds [7]. Context switch is the time required to switch from one process to another process. The queue used in RR is a circular queue [2].

RR is the most prominent scheduling algorithm in time sharing systems. It gives equal priority to each process present in RQ. The response time of processes is reduced into a greater extent. The main objective of RR is to minimize the turn around time and the waiting time, maximize the CPU utilization and reduce the CS [3]. CS is an overhead to the OS.

Scheduling algorithms are divided into two types: preemptive and non-preemptive. In preemptive, higher priority process can preempt the current process in the middle of execution. The current process is moved to RQ. But, Non-preemptive process cannot be released in the middle of execution. A preemptive algorithm may have more CS than Non-preemptive algorithm. RR is a preemptive scheduling algorithm.

Scheduling is done using three schedulers: Long term, Short term and Medium term. Initially, the process is in the spool disk. Long term scheduler is responsible for loading the process from the spool disk to the RQ [2] [4]. When the main memory is free, one of the processes present in the RQ is loaded into main memory. The short term scheduler is responsible for this queuing action. The medium term scheduler is used for I/O execution. The process may require an I/O operation. For I/O execution, the process is moved from the main memory to I/O waiting queue. After I/O operation is over, it is again moved from waiting queue to the ready queue.

CPU scheduling determines which process is allocated to main memory. Each scheduling algorithm is trying to optimize the ATAT, AWT and Average Response Time (ART) [4] [5] [6]. CPU utilization and throughput are also used to measure the performance of an algorithm.

The remaining part of this paperwork is organized as follows. Related work is presented in section 2. The preliminaries are shown in Section 3. Section 4 elaborates the proposed GBTQ RR algorithm with flow chart. The performance analysis is presented in Section 5. We conclude our work in Section 6.

## 2. RELATED WORK
Many researchers have been proposed various methods to improve CPU scheduling. As TQ is inversely proportional to response time, choosing a high TQ will not be wise. Also, static time quantum leads to more CS. So, we have to design such an algorithm which chooses TQ properly as well as CS is very less.

It is better to repeatedly adjust the TQ. Matarneh [6], Panda et al. [1], Bhoi et al. [3] proposes an algorithm based on TQ set. They use a different mathematical measure to choose the TQ. Mostafa et al. [5] uses integer programming to decide the TQ. The TQ is not a too big or too small value. It also reduces the CS.

Noon et al. [4] presents a dynamic TQ mechanism. It also overcomes the demerit of RR such as the fixed TQ problem, CS etc. Bhunia [9] et al. proposes an enhanced version of feedback scheduling. It focuses on the lower priority queue process.

Yuan et al. [10] proposes Fair Round-Robin (FRR). It has a low-complexity scheduler. It gives good short term fairness than STRR [10] [11].

## 3. PRELIMINARIES
Many scheduling algorithms are existing in a multi-programming environment such as FCFS, Shortest Job First (SJF), Shortest Remaining Time First (SRTF), Priority, Highest Response Ratio Next (HRRN), Min-Max Round

Robin (MMRR) [1], Self Adjustment Round Robin (SARR) [6], Virtual Time Round Robin (VTRR) [8], Subcontrary Mean Dynamic Round Robin (SMDRR) [3] etc. The algorithms are listed below. All units are in seconds.

### 3.1 FCFS
It seems like a ticket counter. The process which arrives first in RQ is served first. It is a non-preemptive scheduling algorithm. It means processor cannot release the process before its execution is over. It suffers from Starvation. The process present in the last of RQ has to wait until all process execution is over. So, the TAT and WT are more.

### 3.2 SJF
The algorithm gives priority to shortest process available in RQ. It is also a non-preemptive scheduling algorithm. The process has high Burst Time (BT) suffers most in this algorithm. This type of suffering is called as Aging.

### 3.3 SRTF
It is a preemptive scheduling algorithm. Like SJF, it chooses shortest process first. But, if a new process arrives in RQ, then it compares the new process with the running process. The process which takes less remaining time will occupy the CPU first.

### 3.4 HRRN
The algorithm gives priority to the process which holds the highest response ratio (HRR). The response ratio can be calculated using the equation 1 [2].

$$HRR = \text{Turn around Time} / \text{Response Time} \quad (1)$$

### 3.5 MMRR
It is also a preemptive algorithm. The TQ is repeatedly adjusted in each iteration. It uses Min-Max dispersion measure. The TQ can be calculated using the equation 2 [1].

$$TQ = \text{Maximum Burst Time} - \text{Minimum Burst Time} \quad (2)$$

### 3.6 SARR
Like MMRR, the TQ is adjusted in SARR. It uses Median to repeatedly adjust the TQ. If the TQ is less than 25, then it automatically sets the TQ to 25 [6]. It is also reducing the context switch between processes.

### 3.7 VTRR
It is based on a fair queuing algorithm. O (1) time is required to schedule a client for execution. It was implemented on Linux platform [8].

### 3.8 SMDRR
It uses harmonic mean or subcontrary mean to adjust the TQ. Based on the burst time, it will calculate the harmonic mean. Then, it selects the TQ [3].

## 4. PROPOSED ALGORITHM
### 4.1 Notations

| Notation | Definition |
|---|---|
| RQ | Ready Queue |
| $Q_1$ | First Quartile |
| $Q_2$ | Second Quartile |
| $Q_3$ | Third Quartile |
| N | Total Number of Processes |
| BT [$P_i$] | Burst Time of Process i |
| $RQ_i$ | Ready Queue i |
| TQ [$RQ_i$] | Time Quantum for Ready Queue i |
| MaxBT[$P_k$] | Maximum Burst Time Process k |
| MinBT[$P_l$] | Minimum Burst Time Process l |
| α | Threshold |
| $TQ_{new}$ [$RQ_i$] | New Time Quantum for Ready Queue i |

### 4.2 Descriptions
In our GBTQ algorithm, the processes are sorted in RQ. The quartile measure is used to form a group among the processes. The $Q_1$ is the 25% of the data set. The $Q_2$ (or median) is the 50% of the data set. Finally, the $Q_3$ is the 75% of the data set. It is used in our algorithm because the too short TQ may lead to more CS. Alternatively, the too large TQ may lead to starvation. Based on the CPU BT, the processes are formed four groups. Each group has different TQ. Different TQ is used to reduce CS. As shown in my earlier paper [1], Min-Max dispersion or spread measure was taken to calculate the TQ. The formula is shown in equation 2. It may suffer from CS, if the difference between MaxBT and MinBT is very less. So, in the proposed algorithm, α is used as a threshold to reduce CS. Finally, the TQ is assigned to each group. The process is continued until RQ is empty. After execution of all processes, ATAT, AWT and CS are calculated.

### 4.3 Performance Measure
#### 4.3.1 Turn Around Time (TAT)
It is the overall time a process requires for execution. It can be calculated using the equation 3. The average of all process is termed as ATAT. It can be calculated using equation 4.

$$TAT = \text{Finish Time} - \text{Arrival Time} \quad (3)$$

$$ATAT = \sum_{i=1}^{N} TATi \quad (4)$$

#### 4.3.2 Waiting Time (WT)
It is the queuing delay time require for execution. It may be the time spent in RQ or I/O queue. Normally, it can be calculated using the equation 5. The average WT of all process is termed as AWT. It can be calculated using equation 6.

$$WT = \text{Start Time} - \text{Arrival Time} \quad (5)$$

$$AWT = \sum_{i=0}^{N} WTi \quad (6)$$

#### 4.3.3 Context Switch (CS)
It is the time required to move from one process to another process. In the proposed algorithm, CS is considered as zero. Suppose $P_i$ and $P_j$ are two processes such that $P_j > P_i$. It means $P_j$ is more priority over $P_i$. $T_s(P_i)$ denote the start time of $P_i$ and $T_e(P_j)$ denotes the finish time of $P_j$. Let us assume that $P_i$ immediately follows $P_j$. Then, the CS can be calculated using equation 7.

$$CS = T_e(P_j) - T_s(P_i) \qquad (7)$$

## 4.4 Algorithm

1. Sort the processes present in the RQ.
2. while (RQ != NULL)
3.    Calculate $Q_1$, $Q_2$, $Q_3$.
4.    for i = 1 to N
5.     if BT $[P_i] \leq Q_1$
6.      Place it in $RQ_1$.
7.     else if (BT $[P_i] > Q_1$ && BT$[P_i] \leq Q_2$)
8.      Place it in $RQ_2$.
9.     else if (BT $[P_i] > Q_2$ && BT$[P_i] \leq Q_3$)
10.      Place it in $RQ_3$.
11.     else (BT$[P_i] > Q_3$)
12.      Place it in $RQ_4$.
13.     end if
14.    end for
15.    for i = 1 to 4
16.     Set TQ $[RQ_i]$ = MaxBT$[P_k]$ – MinBT$[P_l]$
17.     if (TQ $[RQ_i] > \alpha$)
18.      Set $TQ_{new}$ $[RQ_i]$ = TQ $[RQ_i]$
19.     else
20.      Set $TQ_{new}$ $[RQ_i]$ = $\alpha$
21.     end if
22.    end for
23.    for i = 1 to N
24.     if ($P_i \in RQ_1$)
25.      $P_i \leftarrow TQ_{new}[RQ_1]$
26.     else if ($P_i \in RQ_2$)
27.      $P_i \leftarrow TQ_{new}[RQ_2]$
28.     else if ($P_i \in RQ_3$)
29.      $P_i \leftarrow TQ_{new}[RQ_3]$
30.     else
31.      $P_i \leftarrow TQ_{new}[RQ_4]$
32.     end if
33.    end for
34. Update N.
35. if (N != NULL)
36.    Go to Step 23.
37. else
38.    Go to Step 40.
39. end if
40. Calculate ATAT, AWT, NCS.
41. end while

**Fig. 1: Proposed Algorithm**

## 4.5 Flow Chart

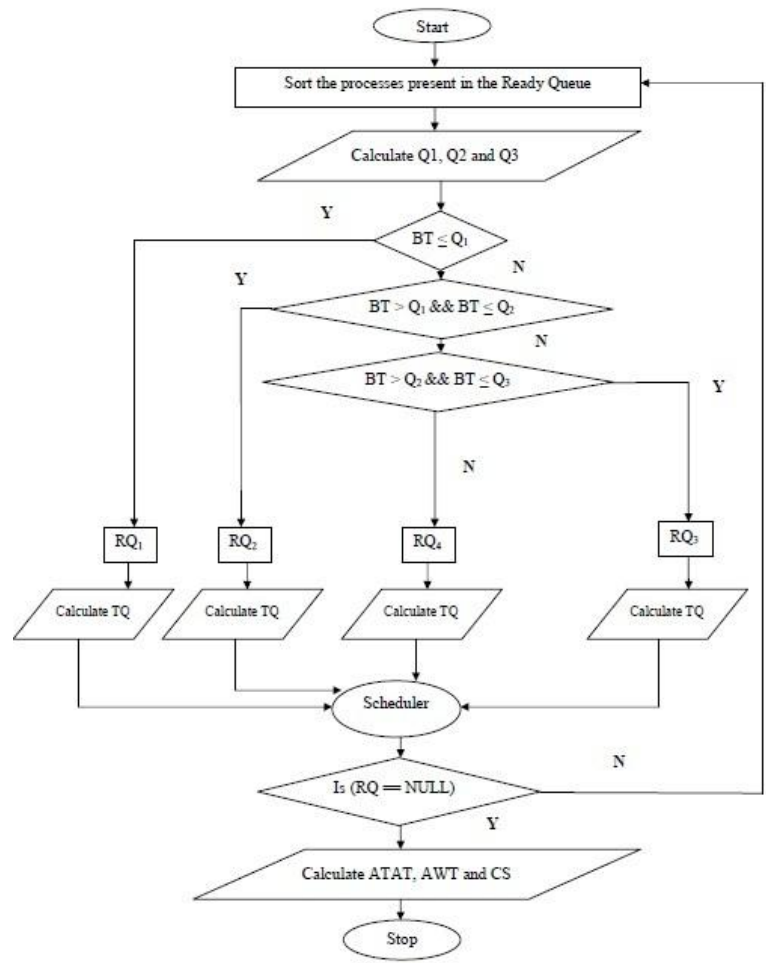

**Fig. 2: Flowchart for GBTQ**

## 5. EXPERIMENTAL RESULTS
### 5.1 Illustrations
We have considered different cases by varying arrival time and burst time. The processes are numbered as $P_1$, $P_2$, $P_3$, ... , $P_N$ where N is the number of processes available in RQ. The below case cover both uniform (U) and non-uniform (NU) BT. The specification of different cases is listed in Table 1.

**Table 1. Case Specifications**

| Case No. | N | Arrival Time | TQ ($\alpha$) | BT Range | U / NU |
|---|---|---|---|---|---|
| 1 | 10 | No | 20 | 7-200 | NU |
| 2 | 4 | No | 20 | 11-95 | NU |
| 3 | 4 | No | 20 | 81-84 | U |
| 4 | 8 | No | 20 | 61-68 | U |
| 5 | 5 | Yes | 20 | 7-75 | NU |
| 6 | 7 | Yes | 20 | 24-150 | NU |

### 5.1.1 Case 1

Let us assume that 10 processes (with AT = 0) have arrived in RQ. The Table 2 shows the AT and BT of each process. In this case, the threshold value is assumed to be α = 20 for processes. The Table 3 shows the comparison of RR and GBTQ respectively. The Figure 3 and 4 shows the gantt chart for RR and GBTQ respectively.

**Table 2. Processes with Burst Time (Case I)**

| Process | Arrival Time | Burst Time |
|---|---|---|
| $P_1$ | 0 | 7 |
| $P_2$ | 0 | 15 |
| $P_3$ | 0 | 24 |
| $P_4$ | 0 | 84 |
| $P_5$ | 0 | 123 |
| $P_6$ | 0 | 145 |
| $P_7$ | 0 | 150 |
| $P_8$ | 0 | 175 |
| $P_9$ | 0 | 180 |
| $P_{10}$ | 0 | 200 |

**Table 3. Comparison of RR and GBTQ (Case I)**

| Algorithm | TQ | ATAT | AWT | CS |
|---|---|---|---|---|
| RR | 20 | 681.3 | 571 | 58 |
| GBTQ | 20,39,30,20 | 610.9 | 498.6 | 44 |

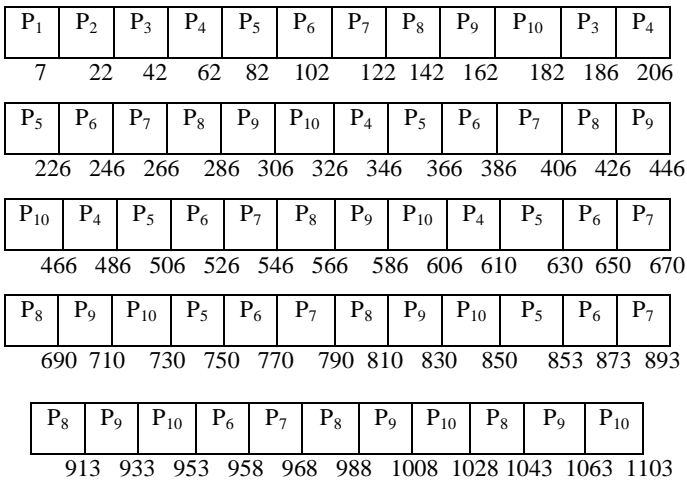

**Fig. 3: Gantt chart for RR**

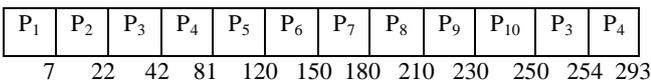

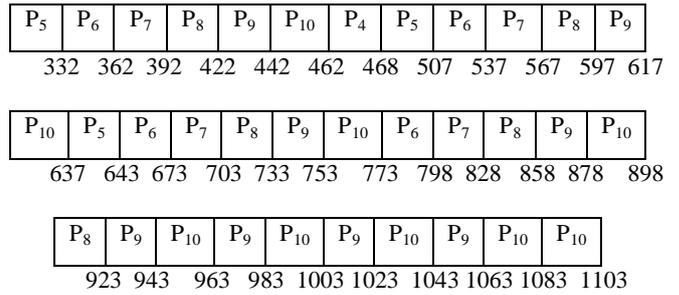

**Fig. 4: Gantt chart for GBTQ**

### 5.1.2 Case 2

Let us assume that 4 processes (with AT = 0) have arrived in RQ. The Table 4 shows the AT and BT of each process. Like Case 1, the threshold is assumed to be α = 20 for processes. The Table 5 shows the comparison of RR and GBTQ respectively. The Figure 5 and 6 shows the gantt chart for RR and GBTQ respectively.

**Table 4. Processes with Burst Time (Case II)**

| Process | Arrival Time | Burst Time |
|---|---|---|
| $P_1$ | 0 | 11 |
| $P_2$ | 0 | 46 |
| $P_3$ | 0 | 82 |
| $P_4$ | 0 | 95 |

**Table 5. Comparison of RR and GBTQ (Case II)**

| Algorithm | TQ | ATAT | AWT | CS |
|---|---|---|---|---|
| RR | 20 | 150.25 | 91.75 | 13 |
| GBTQ | 20, 46, 82, 95 | 110.25 | 51.75 | 3 |

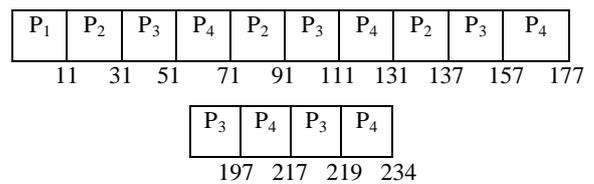

**Fig. 5: Gantt chart for RR**

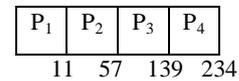

**Fig. 6: Gantt chart for GBTQ**

### 5.1.3 Case 3

Let us assume that 4 processes (with uniform BT) have arrived in RQ. The Table 6 shows the AT and BT of each process. In this case, the threshold value is assumed to be α = 20 for processes. The Table 7 shows the comparison of RR and GBTQ respectively. The Figure 7 and 8 shows the gantt chart for RR and GBTQ respectively.

**Table 6. Processes with Burst Time (Case III)**

| Process | Arrival Time | Burst Time |
|---|---|---|
| $P_1$ | 0 | 81 |
| $P_2$ | 0 | 82 |
| $P_3$ | 0 | 83 |
| $P_4$ | 0 | 84 |

**Table 7. Comparison of RR and GBTQ (Case III)**

| Algorithm | TQ | ATAT | AWT | CS |
|---|---|---|---|---|
| RR | 20 | 325 | 242.5 | 19 |
| GBTQ | 81, 82, 83, 84 | 205 | 122.5 | 3 |

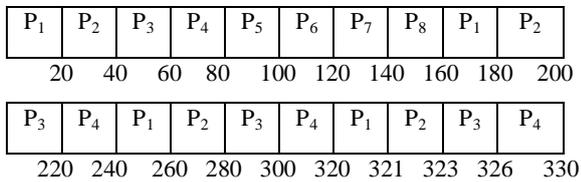

| $P_1$ | $P_2$ | $P_3$ | $P_4$ | $P_5$ | $P_6$ | $P_7$ | $P_8$ | $P_1$ | $P_2$ |
|---|---|---|---|---|---|---|---|---|---|
| 20 | 40 | 60 | 80 | 100 | 120 | 140 | 160 | 180 | 200 |

| $P_3$ | $P_4$ | $P_1$ | $P_2$ | $P_3$ | $P_4$ | $P_1$ | $P_2$ | $P_3$ | $P_4$ |
|---|---|---|---|---|---|---|---|---|---|
| 220 | 240 | 260 | 280 | 300 | 320 | 321 | 323 | 326 | 330 |

**Fig. 7: Gantt chart for RR**

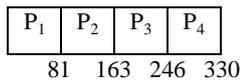

| $P_1$ | $P_2$ | $P_3$ | $P_4$ |
|---|---|---|---|
| 81 | 163 | 246 | 330 |

**Fig. 8: Gantt chart for GBTQ**

### 5.1.4 Case 4

Let us assume that 8 processes (with uniform BT) have arrived in RQ. The Table 8 shows the AT and BT of each process. In this case, assumed value for threshold is α = 20 for processes. The Table 9 shows the comparison of RR and GBTQ respectively. The Figure 9 shows the gantt chart for RR as well as GBTQ.

**Table 8. Processes with Burst Time (Case IV)**

| Process | Arrival Time | Burst Time |
|---|---|---|
| $P_1$ | 0 | 61 |
| $P_2$ | 0 | 62 |
| $P_3$ | 0 | 63 |
| $P_4$ | 0 | 64 |
| $P_5$ | 0 | 65 |
| $P_6$ | 0 | 66 |
| $P_7$ | 0 | 67 |
| $P_8$ | 0 | 68 |

**Table 9. Comparison of RR and GBTQ (Case IV)**

| Algorithm | TQ | ATAT | AWT | CS |
|---|---|---|---|---|
| RR | 20 | 495 | 430.5 | 31 |
| GBTQ | 20, 20, 20, 20 | 495 | 430.5 | 31 |

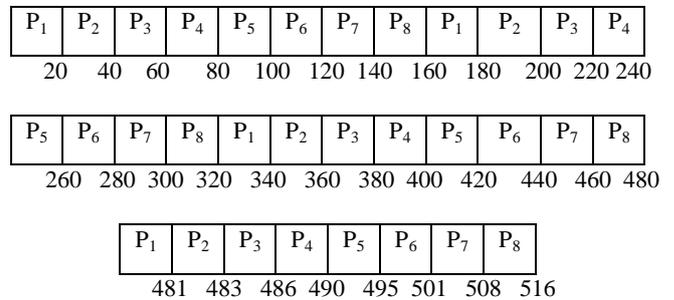

| $P_1$ | $P_2$ | $P_3$ | $P_4$ | $P_5$ | $P_6$ | $P_7$ | $P_8$ | $P_1$ | $P_2$ | $P_3$ | $P_4$ |
|---|---|---|---|---|---|---|---|---|---|---|---|
| 20 | 40 | 60 | 80 | 100 | 120 | 140 | 160 | 180 | 200 | 220 | 240 |

| $P_5$ | $P_6$ | $P_7$ | $P_8$ | $P_1$ | $P_2$ | $P_3$ | $P_4$ | $P_5$ | $P_6$ | $P_7$ | $P_8$ |
|---|---|---|---|---|---|---|---|---|---|---|---|
| 260 | 280 | 300 | 320 | 340 | 360 | 380 | 400 | 420 | 440 | 460 | 480 |

| $P_1$ | $P_2$ | $P_3$ | $P_4$ | $P_5$ | $P_6$ | $P_7$ | $P_8$ |
|---|---|---|---|---|---|---|---|
| 481 | 483 | 486 | 490 | 495 | 501 | 508 | 516 |

**Fig. 9: Gantt chart for RR and GBTQ**

### 5.1.5 Case 5

Let us assume that 5 processes (with AT) have arrived in RQ. The Table 10 shows the AT and BT of each process. In this case, value of threshold is assumed to be α = 20 for processes. The Table 11 shows the comparison of RR and GBTQ respectively. The Figure 10 and 11 shows the gantt chart for RR and GBTQ respectively.

**Table 10. Processes with Burst Time (Case V)**

| Process | Arrival Time | Burst Time |
|---|---|---|
| $P_1$ | 0 | 7 |
| $P_2$ | 5 | 14 |
| $P_3$ | 15 | 55 |
| $P_4$ | 50 | 75 |
| $P_5$ | 75 | 23 |

**Table 11. Comparison of RR and GBTQ (Case V)**

| Algorithm | TQ | ATAT | AWT | CS |
|---|---|---|---|---|
| RR | 20 | 87.4 | 52.6 | 8 |
| GBTQ | 20, 20, 55, 75 | 85.8 | 51 | 4 |

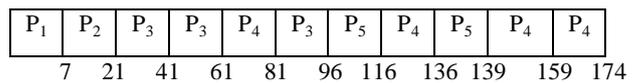

| $P_1$ | $P_2$ | $P_3$ | $P_3$ | $P_4$ | $P_3$ | $P_5$ | $P_4$ | $P_5$ | $P_4$ | $P_4$ |
|---|---|---|---|---|---|---|---|---|---|---|
| 7 | 21 | 41 | 61 | 81 | 96 | 116 | 136 | 139 | 159 | 174 |

**Fig. 10: Gantt chart for RR**

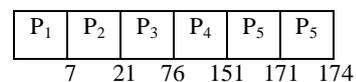

| $P_1$ | $P_2$ | $P_3$ | $P_4$ | $P_5$ | $P_5$ |
|---|---|---|---|---|---|
| 7 | 21 | 76 | 151 | 171 | 174 |

**Fig. 11: Gantt chart for GBTQ**

### 5.1.6 Case 6

Let us assume that 7 processes (with AT) have arrived in RQ. The Table 12 shows the AT and BT of each process. In this case, assumed threshold value is α = 20 for processes. The Table 13 shows the comparison of RR and GBTQ respectively. The Figure 12 and 13 shows the gantt chart for RR and GBTQ respectively.

**Table 12. Processes with Burst Time (Case VI)**

| Process | Arrival Time | Burst Time |
|---|---|---|
| $P_1$ | 0 | 24 |
| $P_2$ | 17 | 48 |
| $P_3$ | 35 | 65 |
| $P_4$ | 50 | 74 |
| $P_5$ | 70 | 89 |
| $P_6$ | 80 | 100 |
| $P_7$ | 130 | 150 |

**Table 13. Comparison of RR and GBTQ (Case VI)**

| Algorithm | TQ | ATAT | AWT | CS |
|---|---|---|---|---|
| RR | 20 | 333.43 | 254.86 | 26 |
| GBTQ | 24, 20, 20, 150 | 327.71 | 249.14 | 25 |

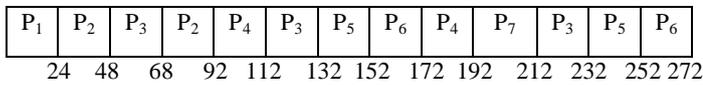
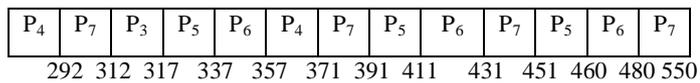

**Fig. 12: Gantt chart for RR**

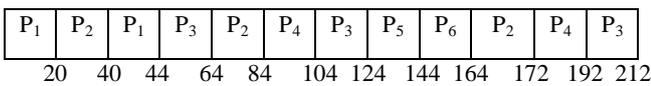
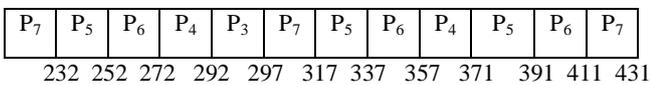
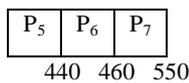

**Fig. 13: Gantt chart for GBTQ**

### 5.2 Experiments

In Section 5.1, six cases are explained. In each case, we compare the proposed GBTQ algorithm with the existing RR algorithm. Both algorithms give the same result in case 4. The experiments show that the proposed algorithm is better than RR algorithm in terms of ATAT, AWT and CS. Performance metrics of different cases are shown in Figure 14, Figure 15 and Figure 16 respectively.

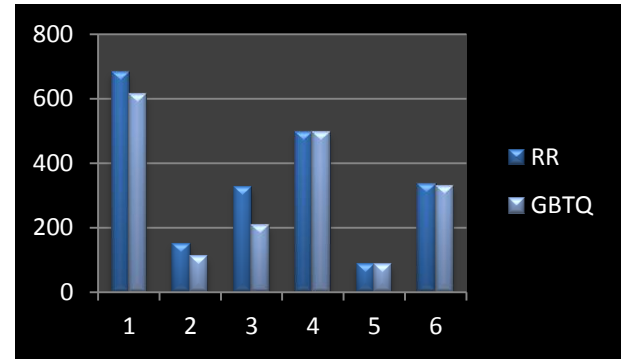

**Fig. 14: Comparison of Turn Around Time**

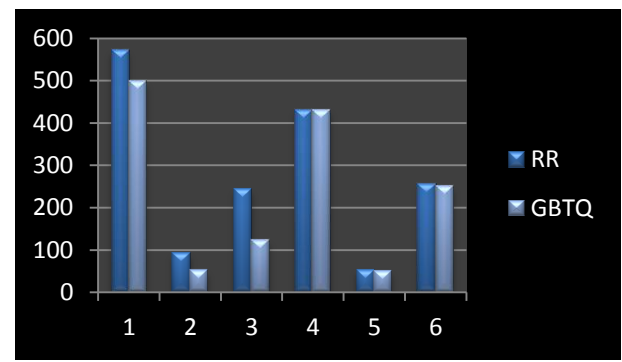

**Fig. 15: Comparison of Waiting Time**

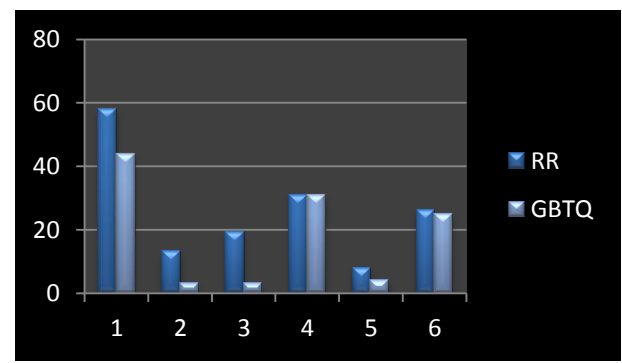

**Fig. 16: Comparison of Context Switch**

## 6. CONCLUSION

Scheduling is a major area in the operating system. Scheduling process may be user process or kernel process. A process may demand for more I/O than the CPU. So, we need an efficient scheduling to compensate CPU process with I/O process. In the proposed GBTQ algorithm, we are focusing on CPU process only. A group based TQ is proposed in this algorithm. Each group has different TQ. This algorithm reduces starvation as well as CS.

In the future, we can extend it to I/O processes. Deadline constraints may be considered as a part of research. We can explore the idea of RR to multi-processor (homogeneous or non-homogeneous) environment.